\documentclass[reprint,amsmath,amssymb,aps,pra]{revtex4-2}

\usepackage{amsthm,mathtools,bm}
\usepackage{booktabs}
\usepackage{xcolor}
\usepackage{enumitem}
\usepackage{array}
\usepackage{float}
\usepackage{slashed}
\usepackage{upgreek}
\usepackage[hidelinks]{hyperref}
\usepackage[nameinlink,noabbrev]{cleveref}

\DeclareMathOperator{\Tot}{Tot}
\DeclareMathOperator{\Tr}{Tr}
\DeclareMathOperator{\Det}{Det}
\DeclareMathOperator{\Td}{Td}
\DeclareMathOperator{\Ch}{Ch}
\DeclareMathOperator{\ad}{ad}

\newcommand{\R}{\mathbb{R}}
\newcommand{\Z}{\mathbb{Z}}
\newcommand{\PTcov}{\widetilde{\mathbb{PT}}_{L}}
\newcommand{\PTorb}{\mathcal{PT}_{L}}
\newcommand{\SU}{\mathrm{SU}}
\newcommand{\G}{\mathsf{G}}
\newcommand{\F}{\mathsf{F}}
\newcommand{\Sc}{\mathsf{S}}
\newcommand{\C}{\mathbb{C}}
\newcommand{\CP}{\mathbb{CP}}
\newcommand{\Zn}{\mathbb{Z}_{n}}
\newcommand{\OO}{\mathcal{O}}

\newcommand{\id}{\mathrm{id}}
\newcommand{\orb}{\mathrm{orb}}
\newcommand{\HK}{\mathrm{HK}}
\newcommand{\dd}{\,\mathrm{d}}
\newcommand{\e}{\mathrm{e}}
\newcommand{\ii}{\mathrm{i}}

\begin{document}

\title{The one-loop QCD scale anomaly on ADE orbifolds as an index}

\author{Herbie Warner}
\affiliation{Rudolf Peierls Centre for Theoretical Physics, University of Oxford\\
Parks Road, Oxford OX1 3PU, United Kingdom}

\begin{abstract}
We derive the one-loop scale anomaly of $\SU(N)$ Yang--Mills on the orbifold $\R^4/\Gamma$, with $\Gamma\subset \SU(2)$ a finite subgroup, from an equivariant index theorem on twistor space. The anomaly separates into the QCD $\beta$-function and a contribution localized at the orbifold fixed point. This localized term inherits the gravitational Weyl anomaly of the smooth ALE resolution, which concentrates under the singular degeneration, along with an additional unprotected heat-kernel coefficient. 
\end{abstract}

\maketitle

\section{Introduction}
Whilst four-dimensional Yang--Mills theory is classically scale invariant, this symmetry is broken by quantum fluctuations. This results in a running of the coupling encoded by the $\beta$-function. Since the original perturbative one-loop calculations \cite{Politzer1973,GrossWilczek1973}, the route has been considerably expedited. The background-field method recast the calculation in a manifestly gauge-covariant form \cite{DeWitt1967,Abbott1982}, while more recently Bittleston and Costello recovered the one-loop $\beta$-function, along with the Weyl-anomaly coefficients, from a short index calculation on twistor space \cite{BittlestonCostello2026}. This belongs to a broader connection between quantum anomalies and index theory \cite{NIELSEN1977493,AlvarezGaumeGinsparg1985}. Herein we extend this approach to ADE orbifolds using equivariant orbifold index theory.

We briefly review the argument of \cite{BittlestonCostello2026} on which our approach is based. An instanton of charge $k$ contributes the classical weight $\exp(-8\pi^2k/g_{\mathrm{YM}}^2)$ to the path integral. Integrating over quantum fluctuations multiplies this weight by determinant factors of the fields, which enter the one-loop instanton measure. Since the complete contribution must be independent of the renormalization scale, the running coupling in the classical weight must compensate the logarithmic scale dependence of this quantum measure. Thus computing this scale anomaly  determines the one-loop $\beta$-function. 

To determine this anomaly, recall that, on twistor space, the fluctuation operators form a Dolbeault complex whose determinant defines a line over instanton moduli space \cite{Quillen1985,BismutFreed1986}. To measure its scale anomaly, one introduces a line bundle $L\to X$ and forms the corresponding family of scaled spacetimes and instanton moduli spaces. The transition functions of $L$ rescale spacetime, while the induced transformations of the determinant line describe how the measure changes. Infinitesimally, a scale weight $A$ contributes $A c_1(L)$ to its first Chern class; a nonzero contribution therefore records the failure of scale invariance \cite{Freed2017}. The family index theorem computes this class, and hence the anomaly. For a charge-$k$ instanton background on $\R^4$, with fundamental matter, \cite{BittlestonCostello2026} obtains
\[
c_1\!\left(\Det^{1/2}\right)=\left(\frac{11}{3}N-\frac{2}{3}N_{\mathrm f}-\frac{1}{6}N_{\mathrm s}\right)k\,c_1(L).
\]
The square root accounts for the inclusion of both fields and antifields in the BV complex \cite{Schwarz_1993}. The coefficient in parentheses is the familiar one-loop $\beta$-function coefficient, which we denote by $\beta$.

We will extend this construction to $\R^4/\Gamma$, with $\Gamma\subset\SU(2)$ a finite subgroup, using Kawasaki's orbifold Riemann--Roch theorem \cite{Kawasaki1979,Kawasaki1981}. Indeed RG flows can be studied on any background admitting a scaling symmetry, making conical geometries natural settings for this construction: for orbifolds dilatations commute with the $\Gamma$-action, so they descend to the quotient, and the RG flow can again be studied through the quantum anomaly of the resulting classical scaling symmetry. 

Kawasaki's formula decomposes the family index into a contribution from the identity element of $\Gamma$ and additional contributions from the fixed loci of the nonidentity elements. Correspondingly, we find
\[
c_{1}\!\left(\Det_{\orb}^{1/2}\right)=\bigl(\beta \kappa+A_{\Gamma}\bigr)c_{1}(L), \qquad \kappa := \frac{k}{|\Gamma|},
\]
where $\kappa$ is the quotient-normalized charge. The identity sector is the ordinary bulk contribution, and therefore reproduces the standard $\beta$-function. By contrast, $A_\Gamma$ is the sum of the nonidentity sectors: since every nonidentity element of $\Gamma$ fixes only the origin, these terms depend only on the equivariant isotropy data of the fields at the orbifold point, rather than on the bulk instanton density. 

We can verify this result using a heat-kernel. For a global quotient, the orbifold projector writes the heat trace as an average over group elements. The identity image reproduces the ordinary bulk heat trace, while each nonidentity image is localized near its fixed locus and produces a logarithmic term supported there \cite{Donnelly1979,DrydenEtAl2008,vonGersdorff2008}. We find that the index-theory and heat-kernel results agree sector by sector. Again as only the origin is fixed, this independently confirms that $A_{\Gamma}$ is localized at the origin.

From the effective-action perspective, the appearance of a localized contribution is entirely expected. The singularity breaks translational invariance, so counterterms supported there are compatible with the symmetries. Moreover, dimensional analysis leaves the defect unit operator as the only marginal local operator. Consequently, $A_{\Gamma}$ renormalizes an independent dimensionless weight associated with the codimension-four orbifold defect, rather than  $g_{\text{YM}}$. Indeed bulk quantum fluctuations on orbifolds are known to generate UV divergences localized on their singular strata \cite{GeorgiGrantHailu2001,vonGersdorff2008,Fursaev1994}.

Since $\mathbb{C}^2/\Gamma$ admits a hyperk\"ahler ALE resolution $X_\Gamma$ \cite{Kronheimer1989ALE}, it is natural to ask how the smooth gravitational scale anomaly is related to the localized term $A_\Gamma$. Although $A_\Gamma$ arises from the nonidentity sectors of the orbifold index theorem, these sectors are not purely gauge-theoretic: their fixed-point denominators are determined by the action of $\Gamma$ on the tangent directions at the singularity. Thus $A_\Gamma$ is sensitive to the local orbifold geometry as well as to the isotropy representations of the quantum fields. Moreover, $\mathbb{C}^2/\Gamma$ is flat on its smooth locus, so any curvature contribution associated with the orbifold geometry is necessarily concentrated at the singularity rather than distributed through the bulk. We show that as the exceptional set of $X_\Gamma$ is collapsed, the integrated Weyl-anomaly contribution does not disappear: its curvature-squared density concentrates at the orbifold singularity while its total integral remains fixed. The resulting localized contribution accounts for one part of $A_\Gamma$. The remaining part is proportional to the superrank of the field complex and arises from a jump in an unprotected heat-kernel coefficient across the singular degeneration. We derive this decomposition index-theoretically; from the heat-kernel perspective, the additional term reflects the fact that the short-time limit and the collapse to the orbifold need not commute \cite{Sher2015,MazzeoRowlett2015}.

\section{Equivariant ADHM on $\mathbf{\R^4/\Z_n}$}

We first consider $\R^4/\Z_n$ (the $A_{n-1}$ singularity). Let $R_i$ denote the one-dimensional representation of $\mathbb{Z}_n$ on which the generator $g$ acts by $\omega^i$, where $\omega=\exp(2\pi\ii/n)$ and all representation indices are understood modulo $n$. Identifying $\mathbb{R}^4\cong\mathbb{C}^2$, the coordinate representation is
\[
  Q:=R_1\oplus R_{-1}, \qquad g:(z_1,z_2)\longmapsto(\omega z_1,\omega^{-1}z_2).
\]

Because an instanton on $\R^4/\Z_n$ lifts to a $\Z_n$-equivariant instanton on the smooth cover, the ADHM vector spaces $V$ and $W$ carry finite-dimensional representations of $\Z_n$. Since $\Z_n$ is finite and we work over $\C$, these representations are completely reducible, so they decompose into their isotypical components \cite{KronheimerNakajima1990,Nakajima1994}. That is
\[
\begin{aligned}
V&=\bigoplus_{i=0}^{n-1}V_i\otimes R_i,&W&=\bigoplus_{i=0}^{n-1}W_i\otimes R_i,\\
v_i&=\dim V_i,&w_i&=\dim W_i,
\end{aligned}
\]
with $\dim W = N$ and $\dim V = k$. The representation carried by $W$ specifies the framing at infinity,
\[
\rho_\infty(g)=\operatorname{diag}\bigl(\id_{w_0},\omega \id_{w_1},\ldots,\omega^{n-1}\id_{w_{n-1}}\bigr).
\]
Consequently, for $\SU(N)$, we require
\[
  \sum_{i=0}^{n-1}i\,w_i\equiv 0 \pmod n \qquad \implies \qquad \det\rho_\infty(g)=1.
\]

The calculation below will be carried out for general $\SU(N)$ equivariant ADHM data. However, for the later comparison with the  purely gravitational contribution of the smooth ALE resolution, we will specialize to the sector with no exceptional gauge flux:  the first Chern class of the gauge bundle along the exceptional cycles vanishes. We therefore need to determine the equivalent condition on the orbifold.

On the smooth cover $\C^2$, the affine equivariant ADHM monad is \cite{KronheimerNakajima1990}
\[
  0\longrightarrow\Lambda^2Q\otimes V\longrightarrow(Q\otimes V)\oplus W\longrightarrow V\longrightarrow 0.
\]
Hence the equivariant $K$-theory class of its middle cohomology is
\[
  [E]=[W]-\bigl(1-Q+\Lambda^2Q\bigr)[V].
\]
Restricting to the fixed point, the origin, gives
\begin{equation}\label{eq: bundle at 0}
  [E|_0]=\sum_{i=0}^{n-1}u_iR_i,\qquad u_i=w_i-2v_i+v_{i-1}+v_{i+1}.
\end{equation}

Now let $\pi:\widetilde X\to \R^4/\Z_n$ be the minimal resolution, and let $\widetilde E$ be the bundle determined by the same equivariant ADHM data. Under the McKay correspondence, the nontrivial representations $R_i$ determine tautological line bundles $\mathcal R_i$ on $\widetilde X$, and \cite{KronheimerNakajima1990}
\[
  c_1(\widetilde E)=\sum_{i=1}^{n-1}u_i\,c_1(\mathcal R_i).
\]
Requiring zero exceptional flux imposes $c_1(\widetilde E)=0;$ thus the condition on the ADHM data becomes $u_i=0$ for all $i\neq0$. Moreover this implies $[E|_0]=N R_0,$ as required. 

\section{The quotient twistor setup}

We use a similar family to \cite{BittlestonCostello2026}, but now take a suitable $\Gamma$-quotient along with enforcing equivariance. Let $X$ be a parameter space for a family of spacetime scales. Instead of choosing a single numerical scale, we package the choice into a complex line bundle $L\to X$, which we assume admits a square root. 

The ordinary twistor space of flat $\R^4$ is the total space of $\OO(1)\oplus\OO(1)\to\CP^1$. To make the spacetime scale vary over $X$, we tensor the two fibre coordinates by $L$. Generalizing the flat-space family of \cite{BittlestonCostello2026}, we define the smooth equivariant covering family
\[
\PTcov:=\Tot\!\left(\OO(1)\otimes L\otimes Q\right)\longrightarrow X\times\CP^1,
\]
such that the $\Z_n$ action is encoded through $Q$. On the base $X\times\CP^1$, set
\[
  q:=c_1(L),\qquad H:=c_1\!\left(\OO(1)\right), \qquad x := H + q.
\]

The group $\Zn$ acts trivially on $X\times\CP^1$ and fibrewise on $\PTcov$ through $Q$. The orbifold twistor family is then the global quotient
\[
\PTorb:=[\PTcov/\Zn]\xrightarrow{\ \pi\ }X.
\]
The space of linearly embedded twistor lines in the covering family is
\[ H^0\!\left(\CP^1,\OO(1)\right)\otimes L\otimes Q\cong L\otimes\C^4.
\]
Taking the quotient therefore gives the family of complexified orbifold spacetimes $[L\otimes\C^4/\Zn]$; after imposing the Euclidean real structure, this is the  family of scaled copies of $\R^4/\Zn$. Hence a local trivialization of $L$ chooses a spacetime scale, and multiplication of that trivialization by $\lambda\in\C^\times$ rescales all spacetime coordinates. 

Let $p:\PTcov\to\PTorb$ be the quotient map and define $K:=K_{\PTorb/X}$, the relative canonical orbifold line bundle. Its pullback is the $\Zn$-equivariant relative canonical bundle on the cover,
\begin{equation}
p^*K=K_{\PTcov/X}=\OO(-4)\otimes L^{-2}.
\label{eq: K on cover}
\end{equation}
We use the same symbol $K$ for this pulled-back equivariant line bundle when working on $\PTcov$. 

Let $\mathcal M^{\orb}_{\kappa,L}\to X$ denote the family of charge $\kappa$ framed $\Zn$-equivariant instanton moduli spaces associated with the fibres of $\PTorb\to X$. On the smooth cover, the Ward correspondence identifies ASD connections with holomorphic bundles on twistor space that are trivial on real twistor lines \cite{Ward1977}. Imposing $\Zn$-equivariance on both sides and then descending by the finite-group action gives the corresponding orbifold statement, while the equivariant ADHM construction supplies the finite-dimensional description of these instanton moduli \cite{KronheimerNakajima1990,Nakajima1994}. Equivalently, $\mathcal M^{\orb}_{\kappa,L}$ parametrizes framed holomorphic orbifold bundles on the twistor fibres whose pullbacks to $\PTcov$ are the $\Zn$-equivariant bundles described above. 

We now construct the equivariant field bundle $\mathcal R$ whose Dolbeault determinant computes the anomaly. Firstly there is a universal orbifold gauge bundle
\[
E\longrightarrow\PTorb\times_X\mathcal M^{\orb}_{\kappa,L},
\]
whose pullback to $\PTcov\times_X\mathcal M^{\orb}_{\kappa,L}$ realizes the universal $\Zn$-equivariant ADHM bundle described above. For matter fields, we will need fractional powers of $K$, so we fix
\[
K^{1/4}=\OO(-1)\otimes L^{-1/2}.
\]
This will encode the twistor homogeneity and scaling lift; we can of course tensor with a flat character line $\mathcal T$ associated to a character $\zeta:\Gamma\to\mu_4$, where $\mu_4$ denotes the group of fourth roots of unity so that $\mathcal T^{\otimes4}\simeq\OO$. However we will just absorb this into the flavour representations.

Gluons are adjoint-valued, with positive- and negative-helicity sectors carrying $\OO$ and $K$, respectively, and therefore contribute $\mathfrak{sl}(E)\otimes(\OO\oplus K)$.

Complex scalars carry $K^{1/2}$. Let $F_{\mathrm s}$ be a unitary $\Gamma$-representation of dimension $N_{\mathrm s}$ (unitary to preserve the action). Then the scalars and their conjugates have internal bundles $E\otimes F_{\mathrm s}$ and $(E\otimes F_{\mathrm s})^\vee$, respectively. Any character associated with a different equivariant lift of the square root can be absorbed into $F_{\mathrm s}$, so the chosen $K^{1/2}$ suffices for both sectors.

Positive- and negative-helicity fermions have geometric factors $K^{1/4}$ and $K^{3/4}$. For $N_{\mathrm f}$ massless Dirac fermions, let $F_{\mathrm f,+}$ and $F_{\mathrm f,-}$ be unitary $\Gamma$-representations of dimension $N_{\mathrm f}$. Again the character factors associated with a different equivariant fourth root can likewise be absorbed into these representations. The two flavour actions may differ, but for simplicity we take the vectorlike choice $F_{\mathrm f,+}=F_{\mathrm f,-}=F_{\mathrm f}$. Both helicities then carry the same internal bundle, consisting
of $E\otimes F_{\mathrm f}$ and its dual.

Writing the scalar and fermion internal bundles as
\begin{equation}
\mathcal W_{\mathrm x}= (E\otimes F_{\mathrm x})\oplus(E\otimes F_{\mathrm x})^\vee,
\end{equation}
with $\mathrm x = \mathrm f$ or $\mathrm s$, the total field bundle is
\begin{equation}
\begin{aligned}
\mathcal R={}&\mathfrak{sl}(E)\otimes(\OO\oplus K)\\
&\oplus\Pi\mathcal W_{\mathrm f}\otimes(K^{1/4}\oplus K^{3/4})\\
&\oplus\mathcal W_{\mathrm s}\otimes K^{1/2},
\end{aligned}\label{eq: field bundle R}
\end{equation}
where $\Pi$ denotes the parity shift. Indeed taking all flavour actions trivial recovers the bundle in \cite{BittlestonCostello2026}. 

We are interested in the square root $\Det_{\orb}^{1/2}$ of the determinant line of the fibrewise Dolbeault complex valued in $\mathcal R$, including its fermionic grading. This line varies over $\mathcal M^{\orb}_{\kappa,L}$. The square root arises because the Dolbeault complex describes the BV fields and antifields, whose functional measure is a semidensity \cite{Schwarz_1993}.

Since $L$ encodes spacetime dilatations, the scale dependence of this measure is recorded by the component of $c_1(\Det_{\orb}^{1/2})$ proportional to $q=c_1(L)$. Equivalently, $q$ is the equivariant parameter for the $\C^\times$ scaling action, and its coefficient gives the infinitesimal scaling weight of the one-loop measure. Our task is therefore to compute this coefficient using the local orbifold family-index formula.

The family $\pi:\PTorb\to X$ is presented by the smooth equivariant family $\PTcov\to X$, and the index formula is expressed in terms of fixed loci on this cover. We state it for finite $\Gamma\subset\SU(2)$. For $g\in\Gamma$, let $\PTcov^{\,g}$ be the fixed subfamily, $N_g$ its normal bundle in $\PTcov$, and
\[
\pi_g:\PTcov^{\,g}\times_X\mathcal M^{\orb}_{\kappa,L}\longrightarrow\mathcal M^{\orb}_{\kappa,L}
\]
the projection. Denote the fibre integral of the local orbifold family-index density by $\mathcal I_{\orb}(\mathcal R)$. Then Kawasaki's Riemann--Roch theorem states
\cite{Kawasaki1979,Kawasaki1981}
\begin{equation}
\begin{aligned}
\mathcal I_{\orb}(\mathcal R)={}&\sum_{[g]\subset\Gamma}\frac{1}{|C_\Gamma(g)|}(\pi_g)_*\\
&\left[\frac{\Td(T_{\PTcov^{\,g}/X})\,\Ch(g;\mathcal R|_{\PTcov^{\,g}})}{\Ch\!\left(g;\lambda_{-1}(N_g^\vee)\right)}\right].
\end{aligned}
\label{eq: kaw orb formula}
\end{equation}
Its degree-two part determines the first Chern class of the determinant line. Thus, for the local scale component,
\begin{equation}
\left.c_1\!\left(\Det_{\orb}^{1/2}\right)\right|_{\mathrm{scale}}=\left.\frac12\left[\mathcal I_{\orb}(\mathcal R)\right]_2\right|_{\mathrm{scale}}.
\label{eq: c1 det 12}
\end{equation}
The factor of $\frac12$ accounts for the BV square root, and $[\cdot]_2$ selects degree two after integration along the fixed fibres. 

In \eqref{eq: kaw orb formula}, the sum runs over conjugacy classes, $C_\Gamma(g)$ is the centralizer of $g$, and $(\pi_g)_*$ denotes integration along the fixed fibres. The Todd class describes their tangent geometry. For an equivariant bundle decomposed into $g$-eigenbundles as $V=\bigoplus_\lambda V_\lambda$, we use the delocalized equivariant Chern character
\[
\Ch(g;V):=\sum_\lambda\lambda\,\Ch(V_\lambda).
\]
In particular, if $V=A\otimes B$ with $g$ acting only on the finite-dimensional factor $A$, then
\[
\Ch(g;A\otimes B)=\Tr(g|A)\,\Ch(B).
\]
The transverse directions enter through the $K$-theoretic Euler class
\begin{equation}
\lambda_{-1}(N_g^\vee):=\sum_{j\geq0}(-1)^j\Lambda^jN_g^\vee.
\label{eq: k theory E class}
\end{equation}
The identity class therefore gives the ordinary family Grothendieck--Riemann--Roch density on the cover, divided by $|\Gamma|$, while the nonidentity classes contribute densities supported on their fixed loci.

For a compact orbifold family, $\mathcal I_{\orb}(\mathcal R)$ represents $\Ch(\operatorname{Ind}_{\orb}(\mathcal R))$. Our twistor fibres are noncompact, so we use only the local scale-anomaly component, integrated with the asymptotic framing fixed. In particular, the identity-sector term is proportional to $qH\Ch_2(E)$, whose spacetime density is the instanton density. A smooth finite-action instanton on the cover extends across infinity in $S^4$ \cite{Uhlenbeck1982}, so its instanton density is integrable. The nonidentity sectors are supported on the compact twistor line over the origin. Thus the scale component of Kawasaki's local index density has a convergent fibre integral with the asymptotic framing fixed.

Finally, integration over the twistor line selects the coefficient of $H$, with $\int_{\CP^1}H=1$. Since we retain only terms linear in $q$, we will truncate to $H^2=q^2=0$.

\section{The scale anomaly on $\mathbf{\R^4/\Z_n}$}

We now apply \eqref{eq: kaw orb formula} to $\Gamma=\Z_n$. Every conjugacy class is a singleton and $C_{\Gamma}(g)=\Gamma$, so each sector carries the common factor $1/n$. The identity element fixes the entire equivariant cover and has zero normal bundle. Thus the identity contribution to $c_1(\Det_{\orb}^{1/2})$ is the covering-space contribution of \cite{BittlestonCostello2026}, multiplied by $1/n$.

For a nonidentity element $g^m$, with $m\neq 0$, neither eigenvalue of $g^m$ on $Q$ is unity. The only fixed vectors in each twistor fibre of $\PTcov$ are therefore the zero vectors, and the fixed subfamily in the cover is the zero section
\[
\PTcov^{\,g^m}=X\times\CP^1\subset\PTcov.
\]
The same statement holds for every nonidentity element in general $\Gamma$. Along this fixed subfamily,
\[
\Td(T\PTcov^{\,g^m}/X)=\Td(T\CP^1)=1+H.
\]

Taking the degree-two component of \eqref{eq: kaw orb formula} and recalling the relation in \eqref{eq: c1 det 12}, Kawasaki's formula gives
\begin{equation}
\begin{aligned}
&c_{1}\!\left(\Det_{\orb}^{1/2}\right)=\frac{1}{2n}\int_{\PTcov}\Bigl[\Td(T\PTcov/X)\Ch(\mathcal R)\Bigr]_{8}\\
  &\quad+\frac{1}{2n}\sum_{m=1}^{n-1}\int_{\CP^{1}}\left[\frac{(1+H)\Ch(g^{m};\mathcal R|_{\PTcov^{\,g^m}})}{\Ch\!\left(g^m;\lambda_{-1}(N_m^{\vee})\right)}\right]_{4}.
\end{aligned}
  \label{eq: kaw expanded}
\end{equation}
Here $[\cdots]_d$ denotes the component of real cohomological degree $d$ before fibre integration.

The first term in \eqref{eq: kaw expanded} is $1/n$ of the covering-space answer, which is just the $\beta$-function \cite{BittlestonCostello2026}. Thus the orbifold action affects only the nonidentity sectors.

For a nonidentity sector, the normal directions to the zero section are just the two fibre directions
\[
  N_m=\OO(1)\otimes L\otimes(R_1\oplus R_{-1}).
\]
Thus dualising gives 
\[
    N_m^\vee = \ell_- \oplus \ell_+,\qquad \ell_{\pm}:=\OO(-1)\otimes L^{-1}\otimes R_{\pm1}.
\]
For a line bundle $\ell$, we have $\lambda_{-1}(\ell)=1-\ell$. Since $\lambda_{-1}$ is multiplicative on direct sums,
\[
  \lambda_{-1}(N_m^\vee)
  =(1-\ell_-)(1-\ell_+).
\]
Both eigenlines have ordinary Chern character $\e^{-x}$, while $g^m$ acts on them with eigenvalues $\omega^{\mp m}$. Thus we obtain
\begin{align}
\Ch\!\left(g^m;\lambda_{-1}(N_m^{\vee})\right)&=\left(1-\omega^{-m}\e^{-x}\right)\left(1-\omega^m\e^{-x}\right)\nonumber\\
&=\left(1-\e^{-x}\right)^2+4s_m^2\e^{-x},\nonumber
\end{align}
where
\[
s_m:=\sin\!\left(\frac{\theta_m}{2}\right), \;c_m := \cos\!\left(\frac{\theta_m}{2}\right), \; \theta_m := \frac{2\pi m}{n}.
\]

We also need the Chern character in the numerator. After decomposing each factor into $g^m$-eigenbundles, the finite-group action enters through the corresponding characters, while the ordinary geometric dependence enters through powers of the canonical bundle $K$. In particular, a power $K^{a}$ contributes the exponential
\[
\exp({ac_{1}\!\left(K|_{X\times\CP^{1}}\right)})=1-4aH-2aq+8a^{2}Hq,
\]
where $y:=c_{1}\!\left(K|_{X\times\CP^{1}}\right)=-4H-2q$ from \eqref{eq: K on cover}. Then
\begin{equation}
  C_{m}(a):=\left[\frac{(1+H)e^{ay}}{\Ch\!\left(g^m;\lambda_{-1}(N_m^{\vee})\right)}\right]_{Hq}=\frac{16s_{m}^{2}\left(a-\tfrac12\right)^{2}-1}
  {8s_{m}^{4}},\label{eq: cm(a)}
\end{equation}
where $[\cdot]_{Hq}$ retains only the coefficient of $Hq$. 

The remaining information is equivariant: the action of $g^m$ on the flavour spaces, on the chosen lifts of the fractional powers of $K$, and on the gauge fibre at the fixed point. Recalling that we included the action on fractional $K$ in the flavour spaces, we write
\[
 \phi_{\mathrm x,m}=\Tr(g^m|F_{\mathrm x}),\qquad  \chi_m=\Tr(g^m|E|_0)=\sum_{i=0}^{n-1}u_i\omega^{mi},
\]
where $\mathrm x=\mathrm f,\mathrm s$. The dual character is $\phi_{\mathrm x,-m}$. Thus the gauge and untwisted matter-double characters are
\begin{align*}
 \Tr(g^m|\operatorname{End}(E))&=\chi_m\chi_{-m},\\
 \Tr(g^m|\mathfrak{sl}(E))&=\chi_m\chi_{-m}-1,\\
 \Tr(g^m|\mathcal W_{\mathrm x})&=\chi_m\phi_{\mathrm x,m}+\chi_{-m}\phi_{\mathrm x,-m}.
\end{align*}

The data $u_i$ can change when the ADHM charge changes. However, adding a regular representation to $V$ while fixing $W$ leaves the fixed-point class unchanged: for any finite $\Gamma\subset\SU(2)$,
\[
 Q\otimes\C[\Gamma]\simeq\C[\Gamma]^{\oplus2},\qquad
 (2-Q)\C[\Gamma]=0.
\]
This changes the covering charge by $|\Gamma|$ while preserving $[E|_0]$ and hence the localized term, with all other equivariant data fixed.

The total internal characters are then
\begin{equation}
 \Xi_m^{\mathrm v}=\chi_m\chi_{-m}-1, \qquad \Xi_m^{\mathrm x}=\chi_m\phi_{\mathrm x,m}+\chi_{-m}\phi_{\mathrm x,-m},
\label{eq: internal char}
\end{equation}
and so we find
\begin{align*}
 \Ch(g^m;\mathcal R_{\mathrm v})&=\Xi_m^{\mathrm v}(1+\e^y), \qquad  \Ch(g^m;\mathcal R_{\mathrm s})=\Xi_m^{\mathrm s}\e^{y/2},\\
 \Ch(g^m;\mathcal R_{\mathrm f})&=-\Xi_m^{\mathrm f}(\e^{y/4}+\e^{3y/4}).
\end{align*}
Using \eqref{eq: cm(a)} with the common BV factor $1/(2n)$ gives
\begin{align}
 A_m^{\mathrm v}&=\frac{\Xi_m^{\mathrm v}}{2n}\bigl(C_m(0)+C_m(1)\bigr)=\frac{4s_m^2-1}{8ns_m^4}\,\Xi_m^{\mathrm v},\label{eq: A gauge}\\
 A_m^{\mathrm f}&=-\frac{\Xi_m^{\mathrm f}}{2n}\bigl(C_m(\tfrac14)+C_m(\tfrac34)\bigr)=\frac{c_m^2}{8ns_m^4}\,\Xi_m^{\mathrm f},\label{eq: A ferm}\\
 A_m^{\mathrm s}&=\frac{\Xi_m^{\mathrm s}}{2n}C_m(\tfrac12)=-\frac{\Xi_m^{\mathrm s}}{16ns_m^4}.\label{eq: A scalar}
\end{align}
Consequently, letting $A_m$ be the sum of above, we find
\[
\left[c_{1}\!\left(\Det_{\R^4/\Z_n}^{1/2}\right)\right]_{q}=\left(\beta\kappa+\sum_{m=1}^{n-1}A_m\right)q.
\]

Later we wish to compare to the gravitational contribution from the smooth resolution. Thus we impose the additional condition $c_1(\widetilde E)=0$, such that  $[E|_0]=NR_0$ and hence $\chi_m=N$. We also take trivial equivariant lifts of the fractional powers of $K$ and trivial orbifold actions on the flavour spaces, so that $\phi_{\mathrm x,m}=N_\mathrm x$. Then using standard trigonometric sums we find with these lifts
\[
A^{\text{can}}_{\Z_n}=\frac{n^{2}-1}{360n}\bigl[(49-n^{2})\G+2(n^{2}-4)\F-(n^{2}+11)\Sc\bigr],
\]
where
\[
    \G=N^2-1, \qquad \F=NN_{\mathrm f}, \qquad \Sc=NN_{\mathrm s}.
\]
Hereafter ``canonical-sector'' denotes zero exceptional first-Chern-class flux, together with these trivial lifts.

\section{The anomaly for ADE-orbifolds}\label{s: ADE res}

The cyclic calculation extends directly to any $\Gamma$ of ADE type. The local geometry entering Kawasaki's formula is unchanged (the only fixed point is still the origin); we just need consider the representation of $\Gamma$. 

Let $\{R_I\}_{I\in\widehat\Gamma}$ be the irreducible representations, with $R_0$ trivial, and decompose the ADHM spaces as
\[
  V=\bigoplus_I V_I\otimes R_I,
  \qquad
  W=\bigoplus_I W_I\otimes R_I.
\]
Writing
\[
  Q\otimes R_I=\bigoplus_J a_{IJ}R_J,
\]
where $a_{IJ}$ is the adjacency matrix of the affine ADE McKay graph, the equivariant ADHM monad gives
\[
  [E|_0]=\sum_Iu_IR_I,\qquad u_I=w_I-2v_I+\sum_Ja_{IJ}v_J.
\]
Again, 0-flux then requires $u_I=0$ for $I\neq0$. 

The identity sector is unaffected by the $\Gamma$ action and therefore just gives $\beta/|\Gamma|$. For $g\neq1$, let the eigenvalues of $g$ on the defining representation $Q$ be $e^{\pm\ii\theta_g}$ and set
\[
s_g:=\sin\!\left(\frac{\theta_g}{2}\right),\qquad c_g:=\cos\!\left(\frac{\theta_g}{2}\right).
\]
The origin is still the only fixed point, so its fixed subfamily in twistor space is again the zero section. Thus the computation leading to \eqref{eq: cm(a)} is unchanged except for replacing $m$ by $g$. Likewise make the same replacement to the internal characters \eqref{eq: internal char}. Then by direct analogy
\[
 A_{[g]}=\frac{2\Xi_g^{\mathrm v}(4s_g^2-1)+2\Xi_g^{\mathrm f}c_g^2-\Xi_g^{\mathrm s}}{16|C_\Gamma(g)|s_g^4},
\]
and so in the canonical sector,
\[
  A_{[g]}^{\text{can}}=\frac{\G(4s_g^2-1)+2\F c_g^2-\Sc}{8|C_\Gamma(g)|s_g^4}.
\]

The conjugacy-class weights may be converted into a uniform group average using
\[
\sum_{[g]}\frac{1}{|C_\Gamma(g)|}f(g)=\frac1{|\Gamma|}\sum_{g\in\Gamma}f(g),
\]
for any class function $f$. Define the two moments
\[
S_2(\Gamma):=\frac1{|\Gamma|}\sum_{g\neq1}\frac1{s_g^2},\qquad S_4(\Gamma):=\frac1{|\Gamma|}\sum_{g\neq1}\frac1{s_g^4}.
\]
Then the complete fixed-point term becomes
\begin{equation}
A_{\Gamma}^{\text{can}}=\frac{2\G-\F}{4}\,S_2(\Gamma)+\frac{-\G+2\F-\Sc}{8}\,S_4(\Gamma).
\label{eq: can ADE A}
\end{equation}
Thus all dependence on the ADE group is reduced to two elementary sums which we give in the Appendix; the field theory enters only through $\G,\F,\Sc$. 

\section{The anomaly via the heat-kernel on $\mathbf{\R^4/\Z_n}$}

The heat-kernel provides a direct spacetime check of the orbifold index result. For a global quotient, the projected heat trace is obtained by averaging the lifted heat kernel over the orbifold group. The identity sector gives $\beta$ \cite{Vassilevich2003}, whereas each $g\neq1$ is controlled, in the short-time limit, by the fixed spacetime locus of $g$ \cite{Donnelly1979,DrydenEtAl2008,vonGersdorff2008}. This is the heat-kernel counterpart of the decomposition appearing in Kawasaki's index theorem. More generally, heat-kernel methods provide one of the standard analytic routes to index densities and their local asymptotic expansions \cite{McKeanSinger1967,Gilkey1995}. Herein we will just use the heat kernel as an independent check of $A_{\Z_n}$.

Whereas recovering the identity-sector logarithm requires evaluating the full $a_4$ Seeley--DeWitt coefficient of the background-dependent gauge, ghost and matter operators, the nonidentity sectors are simpler: their kernels connect $x$ to $g^m x$, and their separation vanishes only at the fixed point. The short-time Gaussian therefore localizes the UV contribution near $x=0$. 

Let $\gamma(g^{m})$ denote the action of $g^{m}$ on all internal indices of a field, and let $(U_{m}\Phi)(x)=\Phi(g^{-m}x).$ For an operator $\mathcal O$ commuting with the orbifold action,
\[
  \Tr_{\orb}\e^{-t\mathcal O}=\frac{1}{n}\sum_{m=0}^{n-1}\Tr\!\left(\gamma(g^{m})U_{m}\e^{-t\mathcal O}\right).
\]
If $K_{\mathcal O}(t;x,y)$ is the integral kernel of $\e^{-t\mathcal O}$, then
\[
\Tr\!\left(\gamma(g^{m})U_{m}\e^{-t\mathcal O}\right)\!=\!\!\int_{\R^{4}}\hspace{-0.3cm}\dd^{4}x\,  \operatorname{tr}_V\!\left[\gamma(g^{m})K_{\mathcal O}(t;g^{-m}x,x)\right],
\]
with $V$ the internal space. For a Laplace-type operator, the short-time expansion near $x=0$ begins with the scalar Gaussian kernel times the identity,
\[
  K_{\mathcal O}(t;g^{-m}x,x)\sim K_{0}(t;g^{-m}x,x)\bigl(\id_{V}+O(t)\bigr),
\]
where the flat kernel is the standard Gaussian
\[
  K_{0}(t;x,y)=\frac{1}{(4\pi t)^{2}}\exp\!\left(-\frac{|x-y|^{2}}{4t}\right).
\]
Since this is already $t^0$ on the nonidentity image integral, the corrections due to connection, endomorphism, and curvature, which enter at positive powers of $t$, do not contribute to the UV logarithm \cite{Gilkey1995}. 

For $m\neq0$, the image integral can be evaluated as
\begin{equation}
I_m(t):=\int_{\R^4}\dd^4x\,K_0(t;x,g^mx)=\frac{1}{16s_m^4},
 \label{eq: I integra}
\end{equation}
as the action rotates one real two-plane by $\theta_m$ and the other by $-\theta_m$. The absence of $t$ implies this is the coefficient of the fixed-point logarithm.

We use the proper-time convention
\[
 \log\Det\Delta=-\int_{\Lambda^{-2}}^{\mu^{-2}}\frac{\dd t}{t}\Tr_{\orb}\e^{-t\Delta}
\]
for the UV part of the determinant. A $t^0$ heat coefficient is therefore multiplied by $2\log(\Lambda/\mu)$. We will find that, for all fields $\Phi$,
\[
A_{m,\HK}^{\Phi}=\frac1n\Xi_m^{\Phi}S_m^{\Phi}I_m,
\]
where $S_m^{\Phi}$ contains the statistics, determinant normalization and geometric character, and $\Xi_m^{\Phi}$ are the internal characters in \eqref{eq: internal char}, which includes the $K$-lift phases too.

For the gauge sector, the background-field operators in Feynman gauge are \cite{Abbott1981,Abbott1982}
\[
 (\Delta_1)_\mu{}^\nu=-D^2\delta_\mu{}^\nu-2\ad(F_\mu{}^\nu),
 \qquad \Delta_0=-D^2.
\]
The real gauge fluctuation gives $\tfrac12\log\Det\Delta_1$ and the complex Faddeev--Popov ghost gives $-\log\Det\Delta_0$. Thus
\[
\begin{aligned}
\Gamma_{\mathrm v}^{\det}&=\frac12\log\Det\Delta_1-\log\Det\Delta_0\\
&=\frac12\int_{\Lambda^{-2}}^{\mu^{-2}}\frac{\dd t}{t}\left[2\Tr_{0,\text{orb}}\e^{-t\Delta_0}-\Tr_{1,\text{orb}}\e^{-t\Delta_1}\right],
\end{aligned}
\]
where the traces are over adjoint-valued zero-forms and one-forms. Since
\[
 \Tr_{T^*\R^4}(g^m)=4\cos\theta_m,
\]
the coefficient of $\log(\Lambda/\mu)$ is given by
\[
 S_m^{\mathrm v}=2-4\cos\theta_m \quad \implies \quad A_{m,\HK}^{\mathrm v}=\frac{\Xi_m^{\mathrm v}(4s_m^2-1)}{8ns_m^4},
\]
in agreement with \eqref{eq: A gauge}.

For matter we use $\mathcal W_{\mathrm x}$ from \eqref{eq: field bundle R}. A complex scalar in $E\otimes F_{\mathrm s}$ has the same Laplacian spectrum as its conjugate, so the doubled trace requires a factor $1/2$:
\[
 \Gamma_{\mathrm s}=\frac12\log\Det_{\mathcal W_{\mathrm s}}\Delta_0=-\frac12\int_{\Lambda^{-2}}^{\mu^{-2}}\frac{\dd t}{t} \Tr_{\mathcal W_{\mathrm s},\orb}\e^{-t\Delta_0}.
\]
The proper-time conversion then gives $S_m^{\mathrm s}=-1$ and
\[
 A_{m,\HK}^{\mathrm s}=-\frac{\Xi_m^{\mathrm s}}{16ns_m^4},
\]
in agreement with \eqref{eq: A scalar}.

For a Dirac fermion, we consider the real part of the effective action. Squaring the first-order operator gives one factor $1/2$, while using the doubled internal bundle gives a further factor $1/2$. Hence
\[
\begin{aligned}
\operatorname{Re}\Gamma_{\mathrm f}&=-\frac14\log\Det_{(S^+\oplus S^-)\otimes\mathcal W_{\mathrm f}} \slashed D^{\,2}\\
&=\frac14\int_{\Lambda^{-2}}^{\mu^{-2}}\frac{\dd t}{t} \Tr_{\mathrm{spin}\otimes\mathcal W_{\mathrm f},\orb}\e^{-t\slashed D^{\,2}}.
\end{aligned}
\]
For the reference geometric spin lift, rotations by $\theta_m$ and $-\theta_m$ give the four spin weights
\[
 \exp\!\left[\frac{\ii}{2}(\alpha\theta_m-\beta\theta_m)\right],\qquad \alpha,\beta=\pm1,
\]
so that
\[
 \Tr_{S^+\oplus S^-}(g^m)=2+2\cos\theta_m=4c_m^2.
\]
The choice $F_{\mathrm f,+}=F_{\mathrm f,-}=F_{\mathrm f}$ makes the internal action identical in both spacetime chiralities. Its character $\Xi_m^{\mathrm f}$ therefore factors from the spin trace. Including the factor $2$ from the proper-time logarithm gives
\[
S_m^{\mathrm f}=2\cdot\frac14\cdot4c_m^2=2c_m^2\qquad \implies\quad A_{m,\HK}^{\mathrm f} =\frac{\Xi_m^{\mathrm f}c_m^2}{8ns_m^4},
\]
which agrees with \eqref{eq: A ferm}.

Thus the heat-kernel and local index calculations agree sector by sector. For any finite $\Gamma\subset\SU(2)$, diagonalising $g\ne1$ on $Q$ gives \eqref{eq: I integra} with $m$ replaced by $g$. 

\section{Relation to the Weyl anomaly}

In this section we compare $A_\Gamma^{\text{can}}$ to the Weyl anomaly of the smooth resolution. For a four-dimensional hyperk\"ahler manifold the one-loop Weyl anomaly is \cite{CapperDuff1974,Duff_1994}
\[
g^{\mu\nu}\langle T_{\mu\nu}\rangle=\frac{1}{(4\pi)^2}\left((c-a)C^2 - \frac{\beta}{2}\operatorname{tr}(F^2)\right),
\]
where $C$ is the Weyl tensor, satisfying $C=-\ast C$ and
\[
a=\frac1{180}\left(31\G+\frac{11}{2}\F+\Sc\right), \quad c=\frac1{60}\left(6\G+3\F+\Sc\right)
\]
are the anomaly coefficients for self-dual QCD \cite{Duff1977}. Then the purely gravitational scale variation can be written as
\[
\delta \log \mathcal{Z}=\frac{-1}{8\pi^2}(c-a)\int_M \operatorname{tr}(C \wedge C)=(c-a)p(M),
\]
where we're interpreting $C$ as a 2-form with values in endomorphisms of $TM$, and $p(M)$ is the Pontryagin number of $M$. Then as the ADE orbifold admits a hyperk\"ahler ALE resolution $X_\Gamma$ \cite{Kronheimer1989ALE}, and $A_\Gamma$ contains geometrical data through the normal bundles, it is natural to wonder how $A_\Gamma^{\text{can}}$ is influenced by the  anomaly on $X_\Gamma$. Indeed the curvature of $X_\Gamma$ is concentrated near the exceptional set, and the integral of curvature squared is invariant under a constant rescaling in four dimensions. It can therefore survive when the exceptional cycles shrink, even though the limiting metric is flat away from the orbifold point.

A simple rearrangement of \eqref{eq: can ADE A} gives
\begin{equation}
A_{\Gamma}^{\text{can}}=6S_2(\Gamma)(a-c)+\left(\frac{S_2(\Gamma)}{30}-\frac{S_4(\Gamma)}{16}\right)\upchi,
\label{eq: A in WA and chi}
\end{equation}
where
\[
\upchi = \operatorname{SRk}(\mathcal{R})=2\G- 4\F + 2\Sc.
\]
We will show below that $6S_2(\Gamma)=-p(X_\Gamma)$, and so the first term in \eqref{eq: A in WA and chi} is exactly the Weyl anomaly. Of course, the appearance of the Weyl anomaly in \eqref{eq: A in WA and chi} does not alone imply it genuinely came from $X_\Gamma$; we will show this is so by considering the limiting support of the curvature as the resolution degenerates. The remaining term will be shown to come from an unprotected heat-kernel coefficient constrained by the hyperk\"ahler nature of $X_\Gamma$.

Hereafter we work with the canonical data: this removes nontrivial gauge, flavour and lift characters, matching the untwisted, purely gravitational smooth calculation. Generally resolutions can carry nontrivial bundles and holonomy, but comparing noncanonical orbifold characters or $K$-lifts requires compatible smooth bundle data and their anomaly contributions as well. 

We first prove $6S_2(\Gamma)=-p(X_\Gamma)$. For an ALE resolution, Chern--Gauss--Bonnet gives \cite{Nakajima1990SelfDuality}
\begin{equation}
\frac{1}{8\pi^2}\int_{X_\Gamma}|Rm|^2 dV_g=\chi(X_\Gamma)-\frac1{|\Gamma|}.
\label{eq: GB result}
\end{equation}
Since our space is also hyperk\"ahler, $Rm=C$, so
\[
p(X_\Gamma)=-2\left(\chi(X_\Gamma)-\frac{1}{|\Gamma|}\right),
\]
where the factor of $2$ follows from the definition of $p$ as a tangent-bundle trace.

For $S_2(\Gamma)$ we note the signature theorem for manifolds with boundary gives the boundary spectral defect as  \cite{Nakajima1990SelfDuality}
\[
\tau=\frac13p(X_\Gamma)-\eta_\Gamma,\quad \eta_\Gamma=\frac1{|\Gamma|}\sum_{g\ne1}\cot^2\!\left(\frac{\theta_g}{2}\right),
\]
where $\eta_\Gamma$ is the signature eta invariant of the asymptotic $S^3/\Gamma$. The exceptional curves have the negative ADE Cartan matrix as their intersection form, so 
\[
\tau=1-\chi(X_\Gamma).
\]
Also, $\cot^2=\csc^2-1$ and there are $|\Gamma|-1$ nonidentity elements so we see
\[
\eta_\Gamma=S_2(\Gamma)-1+\frac1{|\Gamma|} \qquad \implies \qquad 6S_2(\Gamma)=-p(X_\Gamma).
\]
Thus the first term of \eqref{eq: A in WA and chi} is the Weyl anomaly of $X_\Gamma$. 

We now show that the Weyl anomaly survives the degeneration. Fix a hyperk\"ahler ALE metric $g$ on $X_\Gamma$ and set $g_\varepsilon=\varepsilon^2g$. In rescaled ALE coordinates $x=\varepsilon y$, the asymptotic Euclidean metric is fixed while the exceptional set contracts to the origin. Curvature squared scales as $\varepsilon^{-4}$ and the volume form as $\varepsilon^4$, so
\[
|\mathrm{Rm}(g_\varepsilon)|^2\,dV_{g_\varepsilon}
=|\mathrm{Rm}(g)|^2\,dV_g.
\]
Since $a$ and $c$ are fixed by the field content at one loop, the integrated smooth Weyl anomaly therefore remains $(c-a)p(X_\Gamma)$ for every $\varepsilon>0$.

To identify its limiting support, let $r$ be the $g$-distance to the exceptional set; the corresponding $g_\varepsilon$-distance is $\varepsilon r$. For every fixed $\delta>0$,
\[
\int_{\{\varepsilon r>\delta\}}|\mathrm{Rm}(g_\varepsilon)|^2\,dV_{g_\varepsilon}=\int_{\{r>\delta/\varepsilon\}}|\mathrm{Rm}(g)|^2\,dV_g\longrightarrow 0.
\]
Indeed, we have finite total curvature \eqref{eq: GB result}, and the integration region on the right recedes to infinity. Thus every fixed neighbourhood of the collapsing exceptional set captures the entire curvature integral in the limit. Together with \eqref{eq: GB result}, this gives
\[
\frac{1}{8\pi^2}|\mathrm{Rm}(g_\varepsilon)|^2\,dV_{g_\varepsilon}\rightharpoonup\left(\chi(X_\Gamma)-\frac{1}{|\Gamma|}\right)\delta_0,
\]
where $\delta_0$ is the unit point mass at the origin. Thus the Weyl anomaly becomes supported at the orbifold point.

It remains to explore why \eqref{eq: A in WA and chi} has an additional term proportional to $\upchi$ which is absent on $X_\Gamma$. Succinctly on a hyperk\"ahler four-manifold the parallel spinors relate the scalar, fermion and vector kinetic operators, leaving only two independent heat-kernel coefficients.

First, recall that on a hyperkähler 4-manifold the positive-chirality spin bundle is trivial; the global frame is thus given by two linearly independent parallel spinors and so the covariant Laplacian on a section of $S^+$ acts componentwise as the scalar Laplacian on two scalar functions. Writing $D^+:S^+\to S^-$ for the untwisted chiral Dirac operator, $D^-$ for its $L^2$ adjoint,  and
\[
\Delta_+=D^-D^+,\qquad\Delta_-=D^+D^-,
\]
then the Lichnerowicz formula gives $\Delta_+\simeq\Delta_0\oplus\Delta_0$ where $\Delta_0$ is the scalar Laplacian. Moreover as the complexified cotangent bundle is isomorphic to $S^+\otimes S^-$, and $S^+$ is trivial, $\Delta_1\simeq\Delta_-\oplus\Delta_-,$ where $\Delta_1$ is the gauge-fixed one-form Laplacian. The same identities hold on the orbifold since the canonical lift preserves the two parallel spinors.

For either background $B=X_\Gamma$ or $B=\C^2/\Gamma$, let $a_4^B(\Delta)$ denote the integrated coefficient of $t^0$ in the short-time heat-kernel expansion of $\Delta$. On $X_\Gamma$ this means the integral of the local $a_4$ density, whereas on the orbifold, this is the fixed-point contribution. Define
\[
u_B:=a_4^B(\Delta_0),\qquad j_B:=a_4^B(\Delta_+)-a_4^B(\Delta_-).
\]
Using the above identifications between Laplacians
\[
a_4^B(\Delta_-)=2u_B-j_B,\qquad a_4^B(\Delta_1)=4u_B-2j_B.
\]
Then the logarithmic coefficients for a single copy of each matter type are 
\begin{align*}
    A^{\mathrm v}_B&=2a_4(\Delta_0)-a_4(\Delta_1)=-2u_B+2j_B, \\
    A^{\mathrm f}_B&=a_4(\Delta_+)+a_4(\Delta_-)=4u_B-j_B,\\
    A^{\mathrm s}_B&=-2a_4(\Delta_0)=-2u_B,
\end{align*}
and so the total gravitational UV coefficient is
\begin{equation}
A_B:=\G A^{\mathrm v}_B+\F A^{\mathrm f}_B+\Sc A^{\mathrm s}_B=(2\G-\F)j_B-\upchi u_B,
\label{eq: total A on B}
\end{equation}
with $A_{\C^2/\Gamma}=A_\Gamma^{\mathrm{can}}$. Thus the dependence on the field content separates into a chiral local-index contribution and an ordinary scalar heat coefficient. Consequently
\begin{align}
  A_{\mathrm f}(B)+2A_{\mathrm s}(B)&=-j_B,\label{eq: fs j 1}\\
  A_{\mathrm v}(B)+A_{\mathrm f}(B)+A_{\mathrm s}(B)&=j_B.
  \label{eq: fs j 2}
\end{align}
Indeed these are the field contents of an $\mathcal N=2$ hypermultiplet and vector multiplet respectively \cite{ChristensenDuff1979}. Furthermore adding these identities gives 
\[
A_{\mathrm v}(B)+2A_{\mathrm f}(B)+3A_{\mathrm s}(B)=0,
\]
which is the known cancellation of the  gravitational logarithmic coefficient for an $\mathcal N=4$ vector multiplet \cite{Duff_1994}. 

On $X_\Gamma$, $j_{\mathrm{sm}}$ is the bulk local-index contribution associated with the chiral Dirac complex. This is \cite{Gilkey1995}
\[
  j_{\mathrm{sm}}=\int_{X_\Gamma}[\widehat A(TX_\Gamma)]_4=-\frac{p(X_\Gamma)}{24}=\frac14S_2(\Gamma).
\]
We will show $j_{\mathrm{sm}}=j_{\orb}$, following a similar 6 dimensional argument \cite{DegeratuWalpuski2016}, which will allow us to further constrain \eqref{eq: fs j 1}-\eqref{eq: fs j 2} and subsequently derive the $\upchi$ term in \eqref{eq: A in WA and chi}. In particular we use index theory to express each as a chiral Dirac index plus a correction from infinity.  

Let $\rho\geq1$ be a smooth radius function equal to the Euclidean radius $r$ on the end, and use the norms
\[
\|\psi\|_{W^{k,2}_\delta}^2=\sum_{j=0}^{k}\int\rho^{-2\delta-4+2j}|\nabla^j\psi|^2\,dV,\qquad k=0,1,
\]
with $L^2_\delta=W^{0,2}_\delta$. These weighted Sobolev spaces require the spinor and its covariant derivatives through order $k$ to have finite weighted square integrals \cite{DegeratuWalpuski2016}. The parameter $\delta$ controls the permitted behaviour at infinity: a power-law tail $|\psi|\sim r^\alpha$ lies in $L^2_\delta$ precisely when $\alpha<\delta$. The derivative weights account for the additional inverse power of radius contributed by each derivative.

The chiral operators
\[
 D^\pm_\delta: W^{1,2}_\delta(S^\pm) \longrightarrow L^2_{\delta-1}(S^\mp)
\]
are Fredholm for $\delta\in(-3,0)$ \cite{LockhartMcOwen1985,DegeratuWalpuski2016}. The cokernel of $D^+_\delta$ is represented by the kernel of $D^-$ at the complementary weight $-3-\delta$. We henceforth fix $\delta=-3/2$, so these weights coincide and so
\[
 \operatorname{ind}D^+_\delta=\dim\ker D^+_\delta-\dim\ker D^-_\delta.
\]

The ALE Atiyah--Patodi--Singer theorem and the corresponding orbifold formula give \cite{AtiyahPatodiSinger1975I,Kawasaki1981,DegeratuWalpuski2016}
\[ \operatorname{ind}D^+_{\mathrm{sm},\delta}=j_{\mathrm{sm}}-B_D,\qquad\operatorname{ind}D^+_{\orb,\delta}=j_{\orb}-B_D,
\]
where $B_D$ is the boundary spectral correction determined by the asymptotic link $S^3/\Gamma$, its spin structure and the spectral cut selected by the weight. These data agree on both backgrounds, so the same $B_D$ appears in both.

We now show that both weighted indices vanish. On the smooth resolution, weighted elliptic regularity gives
\[
|\psi|=O(r^{-3/2}),\qquad|\nabla\psi|=O(r^{-5/2})
\]
for a spinor in either kernel at weight $\delta=-3/2$. Since $X_\Gamma$ is Ricci-flat, $\slashed D^{\,2}=\nabla^*\nabla$, and so integration by parts on a large truncation $X_R=\{\rho\leq R\}$ yields
\[
\int_{X_R}|\nabla\psi|^2\,dV=\operatorname{Re}\int_{\partial X_R}\langle\psi,\nabla_\nu\psi\rangle\,dS=O(R^{-1})\longrightarrow0,
\]
where $\nu$ is the outward unit normal. Thus $\psi$ is parallel. A parallel spinor has constant pointwise norm, so the imposed decay forces $\psi=0$. Both the kernel and the cokernel therefore vanish, giving $\operatorname{ind}D^+_{\mathrm{sm},\delta}=0 $.

For the orbifold $\C^2/\Gamma$, we use the regular orbifold domain: spinors lift to smooth $\Gamma$-equivariant spinors on $\C^2$. The same Lichnerowicz argument on the flat cover, with the same decay at infinity, shows that both chiral kernels vanish. Hence $\operatorname{ind}D^+_{\orb,\delta}=0$ as well.

Subtracting the two index formulas gives
\begin{equation}
 j_{\orb}=j_{\mathrm{sm}}.
 \label{eq: orb = sm}
\end{equation}
Thus the chiral local-index contribution agrees on the resolution and the orbifold with the asymptotic data fixed.

Alternatively $j_{\orb}$ follows directly from \eqref{eq: I integra}. For the canonical spin lift, the rotations through $\theta_g$ and $-\theta_g$ act on $S^+$ with eigenvalues $1,1$ and on $S^-$ with eigenvalues $e^{\ii\theta_g},e^{-\ii\theta_g}$. The identity image cancels in their difference. Each nonidentity image is independent of $t$, so its spin supertrace is precisely its $t^0$ coefficient:
\[
\begin{aligned}
j_{\orb}&=\frac1{|\Gamma|}\sum_{g\ne1}\bigl(\Tr_{S^+}(g)-\Tr_{S^-}(g)\bigr)I_g(t)\\
 &=\frac1{|\Gamma|}\sum_{g\ne1}\frac{2-2\cos\theta_g}{16s_g^4}=\frac14S_2(\Gamma)=j_{\mathrm{sm}}.
\end{aligned}
\]

Whilst $j$ is conserved, the ordinary scalar heat coefficient $u$ is not index-protected and instead jumps across the singular degeneration. Indeed, define for each species
\[
  d_i:=A_i(\C^2/\Gamma)-A_i(X_\Gamma),
  \qquad i=\mathrm v,\mathrm f,\mathrm s.
\]
Then \eqref{eq: fs j 1}-\eqref{eq: orb = sm} imply
\[
  d_{\mathrm f}+2d_{\mathrm s}=0, \qquad d_{\mathrm v}+d_{\mathrm f}+d_{\mathrm s}=0.
\]
Since the determinants add over field species, the remainder of $A_\Gamma$ after subtracting the smooth Weyl anomaly is
\[
A_\Gamma^{\mathrm{rem}}=\G d_{\mathrm v}+\F d_{\mathrm f}+\Sc d_{\mathrm s}=d_{\mathrm s}(\G-2\F+\Sc)=\frac{d_{\mathrm s}}2\upchi.
\]
This proves structurally that there is only one possible independent remainder, with field weighting $\upchi$. 

Of course the comparison can also be used to reconstruct the smooth anomaly from $A_\Gamma$. The only additional required data is $u_{\mathrm{sm}}$. On a Ricci-flat four-manifold \cite{Vassilevich2003}
\[
u_{\mathrm{sm}}=\frac{1}{180(4\pi)^2}\int_{X_\Gamma}
R_{\mu\nu\rho\sigma}R^{\mu\nu\rho\sigma}\,dV=-\frac{p(X_\Gamma)}{180}.
\]
On the orbifold, \eqref{eq: I integra} gives $u_{\orb}=S_4(\Gamma)/16$. Combining \eqref{eq: total A on B} with \eqref{eq: orb = sm} consequently yields
\[
A_{\C^2/\Gamma}-A_{X_\Gamma}=-\upchi\bigl(u_{\orb}-u_{\mathrm{sm}}\bigr),
\]
and hence
\[
A_{X_\Gamma}=A_\Gamma^{\mathrm{can}}-\left(\frac{S_2(\Gamma)}{30}-\frac{S_4(\Gamma)}{16}\right)\upchi=(c-a)p(X_\Gamma).
\]
Thus the orbifold index, index matching at infinity and one universal scalar heat coefficient reconstruct the gravitational Weyl anomaly, without an explicit ALE metric. 

The origin of this $\upchi$ remainder can be understood from the heat kernel too. For a resolution of size $\varepsilon$, taking $t\to0$ first probes length scales $\sqrt t\ll\varepsilon$ and therefore sees the smooth resolved geometry, whereas taking $\varepsilon\to0$ first at fixed $t$ gives the conical orbifold heat kernel. These two limits need not commute, as is reflected in the joint heat-trace asymptotics of conic degeneration \cite{Sher2015}. Analogous jumps in the $t^0$ heat coefficient occur when smooth planar domains develop corners \cite{MazzeoRowlett2015}, illustrating that convergence of geometries need not imply convergence of their short-time heat coefficients. The $\upchi$ term, not being an index supertrace, is precisely the resulting jump in this coefficient propagated to the vector and fermion sectors by the parallel-spinor identities.

\section{Discussion and conclusion}

We have computed the one-loop scale anomaly on ADE orbifolds using equivariant index theory on twistor space. The anomaly separates into the standard bulk $\beta$-function and a term $A_\Gamma$ supported at the orbifold point. Being localized, $A_\Gamma$ renormalizes the coefficient of the identity operator at the orbifold point, leaving the bulk running unchanged. In the canonical sector, we have shown $A_\Gamma$ inherits the Weyl anomaly of the smooth resolution: as the exceptional cycles collapse, the curvature-squared concentrates at the origin while its total mass remains fixed. The remaining contribution in $A_\Gamma$ is constrained by the hyperk\"ahler structure of $X_\Gamma$ which reduces the scalar, fermion and vector heat coefficients to just two quantities. This, along with index protection in the degeneration, forces the remainder to be proportional to the superrank $\upchi$, with its coefficient determined by the scalar jump. This contribution reflects the noncommutativity of the short-time and singular-collapse limits.

Given we could reconstruct the smooth Weyl anomaly from the orbifold result it is interesting to wonder whether we may recover $a$ and $c$ independently. For conformally coupled matter and vanishing background gauge field, the general gravitational anomaly is \cite{Duff_1994}
\[
\langle T^\mu{}_{\mu}\rangle_{\mathrm{grav}}=\frac{1}{(4\pi)^2}\bigl(cC^2-aE_4+b\nabla^2 R\bigr),
\]
where $E_4$ is the Euler density and $b$ is scheme dependent. Ricci-flatness sets $E_4=C^2$ and $R=0$, so every smooth ALE background used here probes only $c-a$. To circumvent this we could instead consider finite subgroups of $\mathrm U(2)$ which are not contained in $\SU(2)$, so that the resolution need no longer be Ricci-flat. For example the cyclic group $\Gamma=\frac17(1,2)$, generated by
\[
(z_1,z_2)\longmapsto(\zeta z_1,\zeta^2 z_2),\qquad \zeta=\exp(2\pi\ii/7).
\]
The minimal resolution $X$ of $\C^2/\Gamma$ admits scalar-flat K\"ahler ALE metrics \cite{Calderbank_2004}. Its exceptional curves have self-intersections $-4$ and $-2$, so $X$ is also spin. Here $R=0$ but the Ricci tensor does not vanish; thus we obtain a linear combination of $a$ and $c$ independent of the $c-a$ combination obtained above.

One then proceeds similarly: compute $A_\Gamma$ by using Kawasaki's theorem; construct $A_X$ in terms of $c$ and $a$ using characteristic classes; replace the parallel-spinor identities above with K\"ahler identities involving scalar operators twisted by the fractional canonical bundle; and finally match the spin and Dolbeault index contributions with the same asymptotic data to isolate the remaining scalar heat coefficients. As in the ADE calculation, these require an additional universal heat-kernel input, now including the curvature of the twisting line bundles \cite{Vassilevich2003}.

Another interesting extension is to compare noncanonical data with compatible gauge and flavour bundles on the resolution. This would require matching the orbifold data to fluxes through the exceptional cycles and holonomies at infinity, then computing the resulting bundle-curvature contributions to the scale anomaly using twisted index and heat-kernel formulas. Alternatively one could study the joint heat-kernel asymptotics in the transition region $t\sim\varepsilon^2$, where $\varepsilon$ is the resolution scale. This would describe the transition between the smooth and orbifold coefficients and provide input for studying relative determinants under collapse. One could also extend the calculation to orbifolds with positive-dimensional fixed loci. The anomaly would then involve curvature invariants integrated over the singular strata.
\vspace{0.4cm}
\section*{Acknowledgements}
\vspace{-0.5cm}
HW is very grateful to Kevin Costello for suggesting the project and for his insightful guidance throughout. The author would also like to thank Roland Bittleston for helpful conversations. HW is supported by an STFC Doctoral Landscape Award Scholarship in partnership with the Denys Firth Scholarship in Theoretical Physics.

\bibliography{references}
\bibliographystyle{apsrev4-1}

\onecolumngrid
\appendix

\section{ADE group sums and fixed-point coefficients}

\begin{table}[H]
\centering
\renewcommand{\arraystretch}{2.2}
\begin{tabular}{@{}cccc@{}}
\toprule
Orbifold & $|\Gamma|$ & $S_2(\Gamma)$ & $S_4(\Gamma)$ \\
\midrule

$A_n$& $n+1$& $\displaystyle \frac{n(n+2)}{3(n+1)}$& $\displaystyle \frac{n(n+2)(n^2+2n+12)}{45(n+1)}$ \\[2pt]

$D_n$& $4r$& $\displaystyle \frac{4r^2+12r-1}{12r}$& $\displaystyle \frac{(4r^2-1)(4r^2+11)+360r}{180r}$ \\[2pt]

$E_6$& $24$& $\displaystyle \frac{167}{72}$& $\displaystyle \frac{1505}{216}$ \\[2pt]

$E_7$& $48$& $\displaystyle \frac{383}{144}$& $\displaystyle \frac{4529}{432}$ \\[2pt]

$E_8$& $120$& $\displaystyle \frac{1079}{360}$& $\displaystyle \frac{87109}{5400}$ \\

\bottomrule
\end{tabular}
\caption{Group-theoretic moments entering the general ADE fixed-point formula. For the $A$ series we use the convention $A_n\leftrightarrow \mathbb Z_{n+1}$. For the $D$ series we define $r=n-2$, so that $D_n\leftrightarrow \mathrm{Dic}_r$ with $|\Gamma|=4r$. The $E_n$ values are obtained from the character and centralizer data of \cite{LimRota2023}, using $|[g]|/|\Gamma|=1/|C_\Gamma(g)|$ and $2-\upchi_V(g)=4s_g^2$.}
\end{table}

\begin{table}[H]
\centering
\renewcommand{\arraystretch}{1.9}
\begin{tabular}{@{}c l@{}}
\toprule
Orbifold & $A_{\Gamma}^{\text{can}}$ \\
\midrule

$A_n$& $\displaystyle\frac{n(n+2)}{360(n+1)}\Bigl[(48-2n-n^2)\G+2(n^2+2n-3)\F-(n^2+2n+12)\Sc\Bigr]$ \\

$D_n$
& $\displaystyle\frac{1}{1440r}\Bigl[(-16r^4+200r^2+360r-49)\G+(32r^4-40r^2+360r+8)\F-(16r^4+40r^2+360r-11)\Sc\Bigr]$ \\

$E_6$& $\displaystyle\frac1{1728}\left(499\G+2008\F-1505\Sc\right)$ \\

$E_7$& $\displaystyle\frac1{3456}\left(67\G+6760\F-4529\Sc\right)$ \\

$E_8$& $\displaystyle\frac1{43200}\left(-22369\G+141848\F-87109\Sc\right)$ \\

\bottomrule
\end{tabular}
\caption{Fixed-point scale-anomaly coefficients for the ADE orbifolds. In $D_n$, $r:=n-2$.}
\end{table}

\end{document}